\documentclass[%
 reprint,
superscriptaddress,
 amsmath,amssymb,
 aps,
pra,
longbibliography,
]{revtex4-2}

\usepackage{graphicx}
\usepackage{dcolumn}
\usepackage{bm}
\usepackage{physics}

\newcommand\LL{{\scriptscriptstyle \rm L}}
\newcommand\GG{{\scriptscriptstyle \rm G}}
\newcommand\WW{{\scriptscriptstyle \rm Tail}}
\newcommand\TT{{\scriptscriptstyle \rm Center}}
\newcommand\RR{{\scriptscriptstyle \rm R}}

\newcommand\FF{{\scriptscriptstyle \rm F}}

\newcommand\BB{{\scriptscriptstyle \rm B}}

\newcommand\back{{\scriptscriptstyle \rm S}}
\newcommand\source{{\rm s}}
\newcommand\drain{{\rm d}}
\newcommand\MM{{\scriptscriptstyle \rm M}}
\newcommand\interaction{{\rm int}}
\newcommand\doubledot{{\rm dd}}
\newcommand\tunnel{{\rm tun}}
\newcommand\leads{{\rm leads}}
\newcommand\pdagger{{\phantom{\dagger}}}

\begin{document}

\title{Quantum measurement induces a many-body transition\\
	\normalsize
	Backaction channels beyond the ideal detector paradigm 
	\\
	\normalsize 
	cause a transition in the state of the observed system.}

\author{Michael S. Ferguson}
\altaffiliation{These authors contributed equally to this work}
\affiliation{Institute for Theoretical Physics, ETH Zurich, 8093 Zurich, Switzerland}

\author{Leon C. Camenzind}
\altaffiliation{These authors contributed equally to this work}
\affiliation{Department of Physics, University of Basel, 4056 Basel, Switzerland}

\author{Clemens M\"uller}
\affiliation{IBM Quantum, IBM Research Europe - Zurich, 8803 R\"uschlikon, Switzerland}
\affiliation{Institute for Theoretical Physics, ETH Zurich, 8093 Zurich, Switzerland}

\author{Daniel E. F. Biesinger}
\altaffiliation{Current affiliation: Hahn-Schickard, Institute for Information and Microtechnology, Wilhelm-Schickard-Straße  10, 78052 Villingen-Schwenningen, Germany}
\affiliation{Department of Physics, University of Basel, 4056 Basel, Switzerland}

\author{Christian P. Scheller}
\affiliation{Department of Physics, University of Basel, 4056 Basel, Switzerland}

\author{Bernd Braunecker}
\affiliation{SUPA, School of Physics and Astronomy, University of St Andrews, St Andrews KY16 9SS, United Kingdom}

\author{Dominik M. Zumb\"uhl}
\affiliation{Department of Physics, University of Basel, 4056 Basel, Switzerland}

\author{Oded Zilberberg}
\affiliation{Institute for Theoretical Physics, ETH Zurich, 8093 Zurich, Switzerland}

\date{\today}

\begin{abstract}
The current revolution in quantum technologies relies on the ability to isolate, coherently control, and measure the state of quantum systems. 
The act of measurement in quantum mechanics, however, is naturally invasive as the measurement apparatus becomes entangled with the system that it observes. 
Even for ideal detectors, the measurement outcome always leads to a disturbance in the observed system, a phenomenon called quantum measurement backaction. 
Here we report a profound change in the many-body properties of the measured system due to quantum measurements. 
We observe this backaction-induced transition in a mesoscopic double quantum-dot in the Coulomb-blockade regime, where we switch the electron population through measurement with a charge sensor dot. 
Our finding showcases the important changes in behaviour that can arise due to quantum detectors, which are ubiquitous in quantum technologies.
\end{abstract}

\maketitle

Quantum information processing relies on coherently controlling and coupling individual quantum bits (qubits)~\cite{NielsenChuang}. 
There is a wide variety of competing technologies that realize qubits, each with its own advantages and challenges. 
Examples of qubit hardware include electron and nuclear spins in quantum dots~\cite{pla2013high, zajac2018resonantly}, localised charge states~\cite{kim2015microwave}, superconducting devices~\cite{wallraff2004strong, Krantz2019QuantumEngineerGuide}, internal states of trapped ions~\cite{kielpinski2002architecture, home2009complete}, and even Majorana modes in topological materials~\cite{aasen2016milestones}. 
Regardless of the specific realization, a key required component in quantum hardware is a measurement device, i.e., a quantum detector that is used to determine the outcome of computations~\cite{NielsenChuang, wiseman2009quantum}.

A quantum detector couples to the system it measures such that their respective quantum states become correlated during the measurement process. 
In turn, reading out the state of the detector collapses the system towards the outcome of the measurement,
in a process known as backaction~\cite{wiseman2009quantum}.
An ideal detector operates close to the quantum limit~\cite{Gurvitz1996,Korotkov2001b}, where the backaction it imparts is equal to the rate of information gain about the system's state~\cite{Field1993,Elzerman2003,DiCarlo2004,Harbusch2010,Gasparinetti2012,Granger2012,Maradan2014}. 
The interplay between quantum detectors and coherent processes in quantum systems is an active field of research, which touches upon the foundations of quantum mechanics.
Its broad set of achievements include weak values~\cite{Aharonov1988}, quantum feedback circuits \cite{Korotkov2001a,Korotkov2001b}, high-precision amplifiers~\cite{Hosten2008}, quantum state discrimination~\cite{Zilberberg2013}	and stabilization~\cite{Korotkov2001b}, as well as detector-assisted transport~\cite{Zilberberg2014, Bischoff2015}. 
The latter implies that quantum systems, that are coupled to particle-reservoirs, always experience additional many-body energy-exchange channels of backaction beyond those considered in an ideal detector.

Alternatively, Quantum detectors can be thought of as out-of-equilibrium environments whose internal dynamics depends on the measured system. 
Even standard dissipative environments can fundamentally influence quantum many-body systems and their steady-states in often counter-intuitive ways~\cite{Soriente2018, matern2019coherent}. 
Correspondingly, the environment can even trigger abrupt changes in the system's observable properties, i.e., it can induce phase transitions (PTs) or sharp crossovers~\cite{leggett1987dynamics}. 
As a result, new universality classes~\cite{Diehl2010} and novel topological effects~\cite{Diehl2011} emerge in driven-dissipative systems.

\begin{figure*}
	\includegraphics[scale=1]{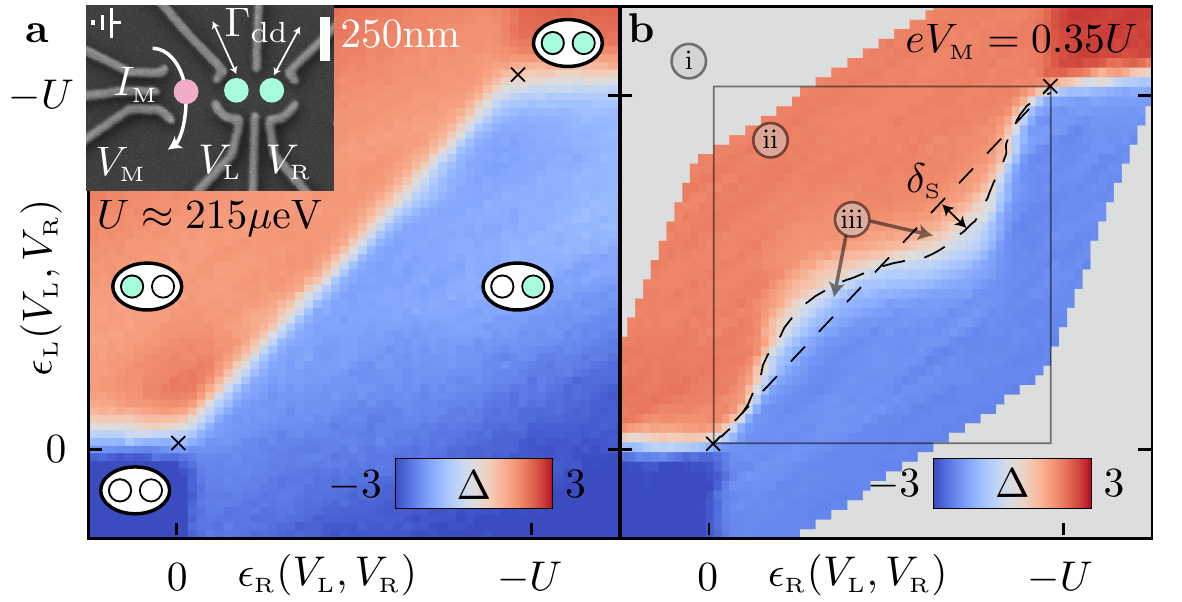}
	\caption{\label{fig:setup}
		\textbf{Charge stability of a double dot with and without the effects of measurement backaction.}
		\textbf{a} DD backacktion imbalance as a function of the dot energies \(\epsilon_\LL\) and \(\epsilon_\RR\) for the case of negligible backaction.
		Since measuring at very low bias is very noisy, we measure at a larger measurement bias \(V_\MM= 150\mathrm{\mu eV}\) and turn on the interdot tunnelling to mimic the canonical DD charge stability diagram as obtained in absence of backaction.
		Full (mint) and empty (white) circles illustrate the ground-state in each region of the observed charge stability map. 
		\textbf{Inset:} Scanning electron micrograph of the device. The population of the DD (mint circles) is monitored by a nearby charge sensor dot (pink circle). 
		\textbf{b}. Same as \textbf{a}, using a measurement bias $V_\MM=75\mathrm{\mu V}$ (\(0.35U/e\)).
		The interdot tunnelling is completely suppressed such that the effects of backaction become apparent in a distortion of the boundary between left and right occupied states. Marked regions (i), (ii), (iii) are governed by different physical mechanisms (see details in the main text).
		Both measurements \textbf{a} and \textbf{b} were performed at an electronic temperature \(T\approx 65\mathrm{mK}\).
	}
\end{figure*}

In this work, we report on a quantum measurement-induced many-body transition. 
We observe this transition in an open many-body quantum system, a double quantum dot (DD) where the two dots are coupled to each other only capacitively, and each dot is tunnel-coupled to a lead. 
An adjacent charge sensor dot (CSD)  measures the charge state of the DD~\cite{Field1993,Elzerman2003,DiCarlo2004,Harbusch2010,Gasparinetti2012,Granger2012,Maradan2014,Biesinger2014}.  
The DD exhibits different phases that are characterized by the charge configuration on the quantum dots. 
In the absence of the detector, these phases are determined by the ground-state energetics of the DD with the leads supplying the charges. 
We observe that the detector imparts backaction that induces a distinct change in the observed charge configuration.
Specifically, we observe that the system now preferentially populates an energetically high-lying state, in close analogy to theoretical predictions for population switching in related systems~\cite{goldstein2010population}. 
We systematically analyse the dependence on measurement strength and temperature of this transition, and develop a concise theory that reproduces the features of the measured data. 
The backaction-induced population switch highlights the extreme sensitivity of quantum systems to out-of-equilibrium fluctuations: For a qubit fully in state \(\ket{0}\), this would correspond to a change into state \(\ket{1}\) by the measurement backaction.

\textit{Experiment -}
We perform our experiments on a gate-defined lateral GaAs device~\cite{Biesinger2014}, composed of a DD (mint) adjacent to a CSD (pink), see inset of Fig.~\ref{fig:setup}\textbf{a}. 
Each single dot in the DD is tunnel-coupled to a separate lead on each side, both with tunnelling rate $\Gamma_\doubledot\sim 100 \mathrm{kHz}$~\cite{supmat}. 
Additionally, Coulomb repulsion between the dots (mutual charging energy) \(U \approx 215 \mathrm{\mu eV}\) imposes an energy penalty on the doubly-occupied states and thus, acts to diminish the occupancy of the DD. 
Crucially to this work, interdot tunnelling is negligibly small in the DD~\cite{supmat}. 
The plunger gate voltages $V_\LL$ and $V_\RR$ are used to tune the left-and right-dot energies $\epsilon_{\LL}$ and $\epsilon_{\RR}$. 
These in turn tune the DD populations $\left(N_\LL, N_\RR\right)$, i.e., the number of electrons of the left and right dots, respectively. 
We perform our experiment in a parameter region, where only four distinct charge states of the DD are relevant, i.e., empty (\(0,0\)), left occupied (\(1,0\)), right occupied (\(0,1\)), or doubly occupied (\(1,1\)). 
In absence of interdot tunneling, the metastability in the canonical charge stability diagram~\cite{Biesinger2014} becomes measurable with the CSD and, simultaneously, all transitions in the DD necessarily involve the reservoirs and are therefore many-body effects~\cite{goldstein2010population}.  
The corresponding charge stability map displays sharp crossovers between distinct DD population configurations, see Fig.~\ref{fig:setup}\textbf{a}.

The population states $\left(N_\LL, N_\RR\right)$ have probabilities \(P_{\left(N_\LL, N_\RR\right)} \) to be observed.
The detector allows us to measure this probability distribution by real-time monitoring of its tunnelling current $I_{\MM}$ in response to an applied bias voltage $V_{\MM}$ across the CSD~\cite{Biesinger2014}. 
This enables direct observation of the population imbalance $\Delta=2P_{\left(1, 0\right)}-2P_{\left(0, 1\right)}+3P_{\left(1,1\right)}-3P_{\left(0, 0\right)}$~\cite{supmat}, which serves as an order parameter for our many-body system.
Microscopically, each charge tunnelling event through the detector realizes a weak measurement kick onto the system through the DD--CSD capacitive interactions. Varying $V_{\MM}$ modifies the current through the detector, thus adjusting the strength of the population measurement~\cite{Gurvitz1996,Korotkov2001b, Zilberberg2014, Bischoff2015}. Applying a small bias voltage, $V_\MM \ll U/e$, results in a conventional charge stability diagram, similar to the one shown in~Fig.~\ref{fig:setup}\textbf{a}, where backaction has no clear effect.

We increase the measurement strength, by increasing the bias-voltage (to $V_\MM=0.35U/e$), and observe a qualitative change in the measured charge stability diagram, see Fig.~\ref{fig:setup}\textbf{b}. 
The diagonal \textit{equilibrium phase boundary} that separates the $(1,0)$ and $(0,1)$ states transforms into an ``S''-shape, with maximal deviation $\delta_{\back}$ from the diagonal. 
In the area enclosed between the ``S'' and the equilibrium phase boundary [region (iii)], the population of the DD is switched, i.e. a high energy state is preferentially occupied.

\begin{figure*}
	\includegraphics[scale=1]{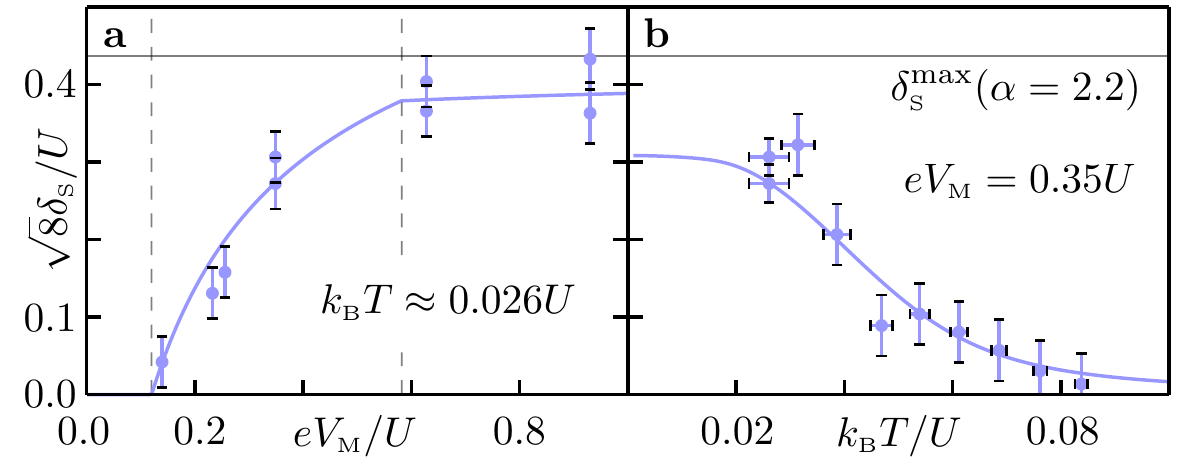}
	\caption{\label{fig:temperature}
		\textbf{Scaling of $\delta_{\back}$.}
		Measured maximal deviation $\delta_{\back}$ (circles) as a function of  ({\bf a}) the measurement bias voltage $V_{\MM}$ at \(k_{{\rm \scriptscriptstyle B}}T/U = 0.026\) (\(65\)mK) and ({\bf b}) temperature $T$ at \(eV_\MM/U=0.35\) (\(75\mathrm{\mu eV}\)).
		Error bars include the uncertainty in determination of the charge-degeneracy curve at $\Delta = 0$ and the uncertainty in the relationship between the fridge temperature and electronic temperature. 
		The theoretical model (cf.~Fig.~\ref{fig:model}), with backaction induced Lorentzian broadening, reproduces the observed temperature and measurement bias dependence of the effect.
		The horizontal solid line indicates the maximal possible deviation~$\delta_{\back}$ predicted by our model for \(\alpha=2.2\), while in the infinite \(\alpha\) limit the saturation is limited by the mutual repulsion and a geometric factor  \(\delta_\back^\mathrm{max}(\alpha\to\infty) = U/\sqrt{8}\)~\cite{supmat}.
		Vertical dashed lines mark points that characterise the two fit parameters \(\chi\) and \(\xi\) of our model.
	}
\end{figure*}

We systematically study the size of the population switching area as a function of applied measurement voltage $V_\MM$ and electronic temperature $T$, see Fig.~\ref{fig:temperature}.
At low temperature, we find that the ``S''-shape grows, and finally saturates with the bias-voltage \(V_\MM\).
In contrast, an increase in temperature washes out the effect.
We conclude that at low temperatures and large bias, backaction dominates the behaviour of the DD.
This is the main result of this work, and highlights how sensitive the state of a quantum system can be on the measurement strength.
Specifically, the quantum measurement induces an abrupt switch in the state of the DD and its leads.
As the DD is coupled to large leads that absorb and emit particles through the transition and there is no interdot tunnelling, this is necessarily a many-body effect~\cite{goldstein2010population}.
Furthermore, we extract the width of the transition in the charge stability diagram~\cite{supmat}.
We observe that even though the amplitude of the ``S''-feature is determined by the backaction strength, the width of this curve, remarkably, is independent of the sensor bias. 
Instead it  reflects the thermal broadening of the reservoirs~\cite{DiCarlo2004,supmat}. 

\begin{figure}
	\includegraphics[scale=1]{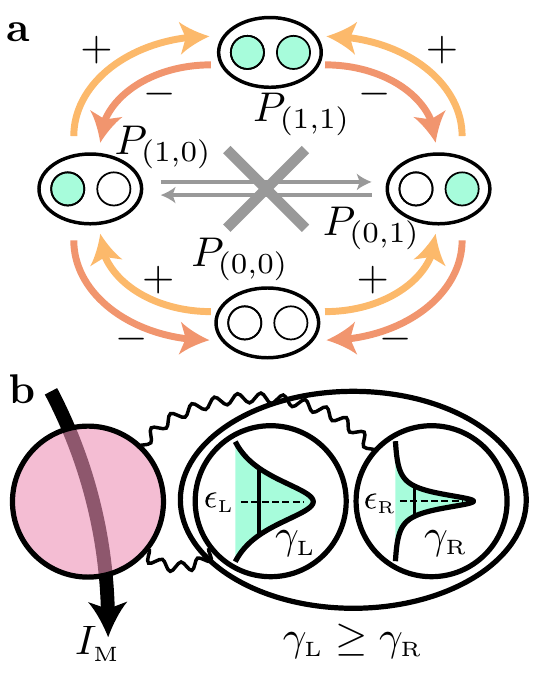}
	\caption{\label{fig:model}
		\textbf{Effective model.}
		\textbf{a}.
		Illustration of the rate equation describing the system dynamics (DD coupled to its leads).
		Sequential tunnelling rates raise (orange arrows) or lower (red arrows) the DD's population by a single electron.
		Direct or virtual left-to-right charge tunnelling is negligible in our system (crossed-out gray arrows)~\cite{supmat}.
		\textbf{b}. The measurement backaction from the nearby charge sensor dot (pink circle) effectively imparts a different width $\gamma_i$ to each dot, $i=L,R$.
	}
\end{figure}

\textit{Theoretical model -}
We develop a concise theoretical model, which provides an intuitive picture of the processes at stake, and reproduces the key features of the experiment.
The open many-body dynamics of the system are effectively described using a rate equation, $\partial_t\mathbf{P}=\Gamma \mathbf{P}$, where $\mathbf{P}$ is a vector of the charge configuration probabilities, and $\Gamma$ is a matrix of transition rates between charge configurations~\cite{Bruus2004, Biesinger2014}. 
In our case, each transition between the DD charge states involves a lead to which a dot is tunnel-coupled. 
To lowest order, transitions between charge states of the DD 
occur through single electrons that hop between the DD and the leads~\cite{Biesinger2014}, 
see Fig.~\ref{fig:model}\textbf{a}. 
We thus neglect direct $(1,0)\leftrightarrow (0,1)$ transitions and cotunnelling between the left and right dots~\cite{supmat}. 
Hence, left-to-right \textit{switching rates} involve either the $(0,0)$ or $(1,1)$ and a motion of charges in the DD necessarily involves the leads, i.e. such transitions are many-body effects.
Without detector backaction, the transition rate $\Gamma_{if}^{ \pm}$ from an initial ($i$) to a final ($f$) charge configuration is $\Gamma_{if}^\pm = \Gamma_{\doubledot}n_\FF(\epsilon_f-\epsilon_i)$, where $+$ or $-$ mark raising or lowering the number of electrons in the DD, respectively. 
Here, we introduced the Fermi-Dirac distribution $n_\FF$, and the energies $\epsilon_{i / f}$ of the initial/final state~\cite{supmat}. 
We note that the Fermi-Dirac distribution of the electrons in the DD leads is the only place where the temperature enters our model.

The motion of charges through the detector capacitively modulates the energy levels in the DD. 
Using a quantum mechanical analysis of the transport in the system given such an out-of-equilibrium measurement, some of us have found that the detector imparts backaction onto the DD in the form of an effective broadening of its levels~\cite{Zilberberg2014}. 
The Lorentzian broadening is proportional to the measurement strength and thus to $V_{\MM}$, and we write it as $\gamma_{\back, \LL} = \alpha \chi  V_{\MM}$ for the left dot and $\gamma_{\back, \RR} = \chi  V_{\MM} / \alpha$ for the right dot. 
The dimensionless fit parameter \(\chi\) depends on the microscopic details of the detector-DD interaction~\cite{supmat, Zilberberg2014}.
Since the distance between the detector and each of the two dots is not equal, the backaction-induced broadening is different for each dot, see Fig.~\ref{fig:model}\textbf{b}. 
We quantify this asymmetry through a parameter $\alpha=2.2\pm0.2$, which we independently extract from the measurement data~\cite{supmat}. 
Note, that the asymmetry \(\alpha\) is necessary to enable the charge sensor dot current to differentiate between the left and righ occupied states.
Other competing environmental effects, such as charge noise, induce an equal broadening on each level on the order of \(\sim 1 \mathrm{\mu eV}\)~\cite{Camenzind2018HyperfinephononSpinRelaxation}.
We include these effects through a fitting parameter \(\xi\) that encodes the width and assume that the total width of the DD levels to be equal to its largest contribution, such that \(\gamma_{j} = \max(\xi, \gamma_{\back, j})$ with $j={\rm L},{\rm R}\).

The level broadening $\gamma_{j}$ modifies the system's transitions rates to $\Gamma_{if}^{\pm} = \Gamma_{\doubledot}n_\LL(\epsilon_f-\epsilon_i,\gamma_j)$, where $n_\LL(\epsilon,\gamma)$ 
is a broadened Fermi function~\cite{supmat}.
Using these modified rates, we compute the steady-state population imbalance~$\Delta$, and extract the maximal deviation~$\delta_\back$ as a function of $V_\MM$. The theoretical results in Fig.~\ref{fig:temperature}, are obtained with fit parameters~\(\chi = 9.4\times10^{-3}\) and~\(\xi= 2.5\times 10^{-3}U \)~(\(0.55 \mathrm{\mu eV}\)).

\begin{figure}
	\includegraphics[scale=1]{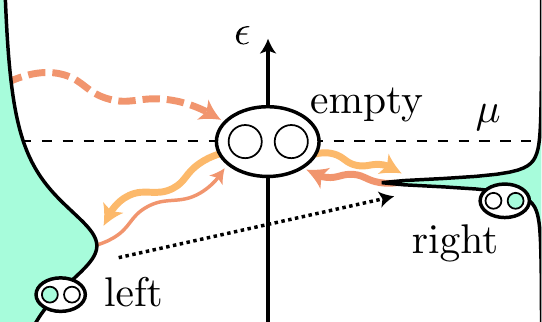}
	\caption{\label{fig:theoryUnderstanding}
		\textbf{Population switching due to an asymmetric detector-induced broadening.}
		The thermal transition rates (solid) into (yellow) both levels are roughly equivalent as both are well below the chemical potential~\(\mu\).
		On the other hand, the thermal rates (solid) out  (orange) of the right dot is much larger, as it is closer to the chemical potential.
		This imbalance, in the absence of level broadening, would lead to the left level being preferentially occupied.
		However, the broad tail of the left level provides an additional rate (dashed) out of the left level that causes a switch in the population, i.e. overall the probability flows from the left to right occupied state (dotted arrow).
		Such a configuration is found in the lower left switching region of Fig.~\ref{fig:setup}\textbf{b}.
	}
\end{figure}

\textit{Physical interpretation -}
The appearance of the the ``S''-shaped charge stability diagram is a direct result of the measurement-induced imbalance $\alpha$ between the broadening of each dot, i.e., due to the stronger coupling between the dot closer to the sensor compared to the more distant one. 
To better illustrate the details of this many-body effect, we divide the transition rates as $\Gamma = \Gamma^\WW+\Gamma^\TT$, where $\Gamma^\TT$ are standard thermally-activated rates arising from the centre of the level's spectral weight distribution, see Fig.~\ref{fig:theoryUnderstanding}.
Thermal contributions to the effective broadening decrease equally for the left and right dots as they are detuned from the chemical potential.
The backaction, on the other hand, broadens the tails of the levels by a different amount for the left and right dots, leading to a more rapid reduction of the tails in the right dot when compared to the left one.
In the tails ($\Gamma^\WW$) the backaction dominates over thermal effects, and the DD population can be controlled by the \emph{difference} between left and right dot level broadening. 
Specifically, when the dot levels are far detuned from one another, or are close to the leads' chemical potentials [regions (i) and most of (ii) in Fig.~\ref{fig:setup}\textbf{b}], the thermal parts dominate the rates and the tails are unimportant. 
Conversely, when the dots' levels are nearly degenerate and far from the chemical potentials [region (iii) and (ii) close to the transition in Fig.~\ref{fig:setup}\textbf{b}], the tails dominate the occupation probability, leading to a switch in the population, with a high-energy state preferentially occupied, see Fig.~\ref{fig:theoryUnderstanding}. 

Our theoretical treatment is able capture the core of the measurement backaction effect, and remarkably, it already quantitatively reproduces the scaling behaviour of the observed phenomena. 
We find that the exact shape of the level broadening, is not crucial to describe the backaction induced population switching dependence in Fig~\ref{fig:temperature}~\cite{supmat}. 
On the other hand, the exact form of the ``S''-shape that our model produces, depends strongly on the exact nature of the tails of the distribution~\cite{supmat}. 
Our experimental and theoretical results imply that the backaction-induced asymmetry in the level width induces the population switching.
Simultaneously, we find that the type of broadening controls the exact phase boundaries and the width of the transitions. 
Our results suggest a sharper than Lorentzian broadening, which could be due to, for example, either energy-dependent widths or higher-order charge correlations in the CSD~\cite{supmat}.
Our results thus leave room and motivate  further refinement through inclusion of more microscopic details of the detector.

\textit{Conclusions -}
Changing the nature of a many-body state simply by observing it is a major shift in how we understand and employ the act of measurement in quantum mechanics. 
The detector backaction broadens the system's energy levels, leading to a complete switch in the system's electron populations.
This is well beyond the paradigm of ideal detectors and highlights the difficulty of keeping a system isolated yet still measurable.
Our results are applicable to a wide variety of experimental systems, ranging from quantum dots to superconducting systems and to photonic microcavities. 
We therefore expect similar effects to manifest themselves in a wide range of quantum information processing hardware. 
\\
\textit{Acknowledgments.}
We acknowledge J. D. Zimmerman and A. C. Gossard for the growth of the GaAs heterostructure.
\\
\textit{Data availability.}
The data supporting this study are available in a Zenodo repository~\cite{Ferguson2020BackactionZenodo}.
\\
\textit{Funding.} Work in Basel was supported by the Swiss Nanoscience Institute,  NCCR  QSIT and SPIN, Swiss NSF Grant No. 179024, ERC Starting Grant (DMZ), and the EU H2020 European Microkelvin Platform EMP, Grant No. 824109. C.P.S. further acknowledges support by the Georg H. Endress Foundation. The authors at ETH and IBM acknowledge financial support from the Swiss National Science Fund directly, and through NCCR QSIT.
\\
\textit{Author contributions.}
LCC, DEFB, CPS and DMZ performed the experimental work. 
MSF, CM, BB, and OZ performed the theoretical work. 
MSF, LCC, CM, and OZ wrote the manuscript. 
All authors discussed the results and the manuscript.

%

\newpage
\cleardoublepage
\renewcommand{\figurename}{Supplementary Figure}
\renewcommand{\appendixname}{ }
\appendix

\setcounter{equation}{0}
\setcounter{figure}{0}
\setcounter{table}{0}

\begin{center}
	\textbf{\Large Supplementary Information}
\end{center}
\section{Experimental setup}
The measurements are performed in a surface-gate defined double quantum dot device fabricated on top of a GaAs two-dimensional electron gas (2DEG) with a nominal density $n=2.6 \cdot 10^{11} \mathrm{cm}^{-2}$ and a mobility $\mu=4 {\cdot} 10^5 \mathrm{cm^2/Vs}$. The device is cooled down to a base temperature of $25$mK in a $^3$He/$^4$He dilution refrigerator equipped with home-built microwave filters resulting in a sample electron temperature $T \approx 65$mK~\cite{Scheller2014}.
Applying negative voltages to surface gates allows us to locally deplete the underlying 2DEG and form two quantum dots in the centre of the device, see inset of Fig.~\ref{fig:setup}\textbf{a}. Each quantum dot is tunnel-coupled to its respective lead with coupling-rates \(\Gamma_\doubledot\) on the order of  \(100\)kHz, (see Supplementary Fig.~\ref{Fig:S4}), while the interdot coupling is reduced to a few Hz. For such small coupling strengths, direct (left \(\leftrightarrow\) right) tunneling and cotunneling processes through the reservoirs are strongly suppressed.
We thus discard cotunnelling and direct tunnelling (crossed out arrows in Fig.~\ref{fig:model}\textbf{a}) as these rates are very small.
We can enable direct tunnelling to wash out the asymmetric effect of backaction. 
This is how we obtained the reference charge stability diagram of Fig.~\ref{fig:setup}, where the interdot coupling washes out the backaction effect, despite a relatively large measurement-bias~\(V_\MM = 150\mathrm{\mu V}\) (\(0.70 U/e\)).
Note that, in comparison to similar experiments, we generally apply a relatively small sensor-bias \(V_\MM\), and $e V_M$ is well below the orbital energies of the DD and CSD.

\begin{figure*}
	\centering
	\includegraphics[scale=1]{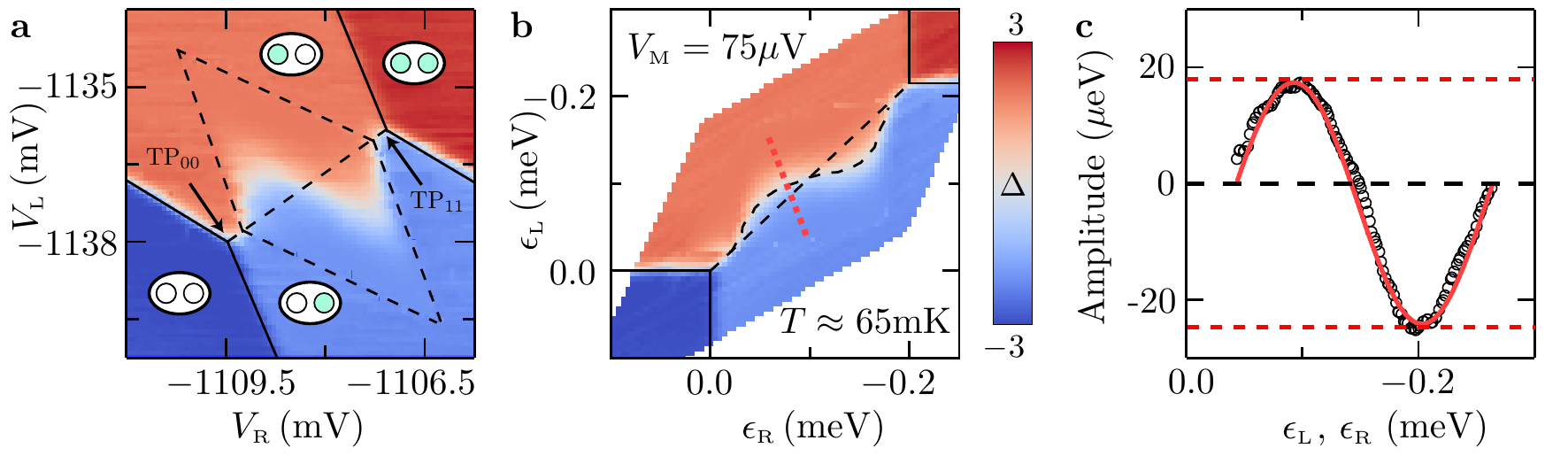}
	\caption{\textbf{Extracting the amplitude of the ``S"-shape.} 
		Measured charge state probabilities around the (1,0)-(0,1) charge degeneracy. 
		\textbf{a}. A scan as a function of the left plunger gate voltage $V_\LL$ versus the voltage on the right plunger gate $V_\RR$. Probabilities inside the dashed diamond were obtained by analysing real-time data traces of sensor conductance~\cite{Biesinger2014}. 
		For the data outside the diamond, the sensor conductance was mapped to charge state configurations. 
		The resulting occupation imbalance $\Delta$ maps are identical but the real-time measurements give more information such as the switching rate (see Supplementary Fig.~\ref{Fig:S4}).
		The zero-detuning line \(\epsilon_\LL=\epsilon_\RR\) is emphasized as a dashed line stretching between the triple-points (TPs).
		The latter, labelled by TP$_{00}$ and TP$_{11}$, mark degeneracy between three possible charge configurations. 
		\textbf{b}. The same data rotated into the $\epsilon_\LL$-$\epsilon_\RR$ basis, i.e., as a function of the energy of the left and right quantum dots, respectively. 
		As the electron has a negative charge, positive voltages correspond to negative energies. 
		The red dotted line, indicates a cut, perpendicular to the transition line, used to compute the transition widths in Supplementary Fig.~\ref{Fig:S4width}.
		In total for Supplementary Fig.~\ref{Fig:S4width} we use four cuts distributed along the length of the transition line.
		\textbf{c}. Extracted ``S''-shape where its maximal amplitude $\delta_\back$ is obtained by fitting a sine model to this data (red curve). 
		The pronounced amplitude offset is discussed in Supplementary Information~\ref{SupNoteExp:4}.}
	\label{Fig:S1}
\end{figure*}

The charge state of the double quantum dot (DD) is continuously monitored by the capacitively-coupled sensor quantum dot operating as a charge sensor dot (CSD) on the left side of the device, see inset Fig.~\ref{fig:setup}\textbf{a}. The charge sensor bandwidth of about \(15\)kHz is limited by the capacitance of the low-pass microwave filters~\cite{Scheller2014} on the input of the current-to-voltage converter (Basel Precision Instruments LNHS LSK389A).
We tune the double dot close to the (1,0)-(0,1) charge degeneracy, see~Fig.~\ref{fig:setup}\textbf{a}. 
Here, due to a low inter-dot tunnel-coupling  compared with the coupling to the leads, a diamond shaped region bounded by the extension of the lead--dot transitions appears, where metastable (1,0)$\leftrightarrow$(0,1) charge state switching occurs~\cite{Biesinger2014}. 
At each position within the diamond, the charge state switching is recorded in real-time over a large number of switching events and digitized ~\cite{Biesinger2014}.
From such real-time traces, the average state-occupation probabilities (Supplementary Fig.~\ref{Fig:S1}),
as well as the switching frequency and associated transition rates (Supplementary Fig.~\ref{Fig:S4}) are calculated from the accumulated times spent in each level~\cite{Biesinger2014}. 
We then use the real-time data to map the conductance of the CSD to the population imbalance \(\Delta\).
In turn, this allows us to extract the amplitude of \(\delta_\back\) over a large set of parameters \(T,V_\MM\).

\begin{figure*}
	\centering
	\includegraphics[scale=1]{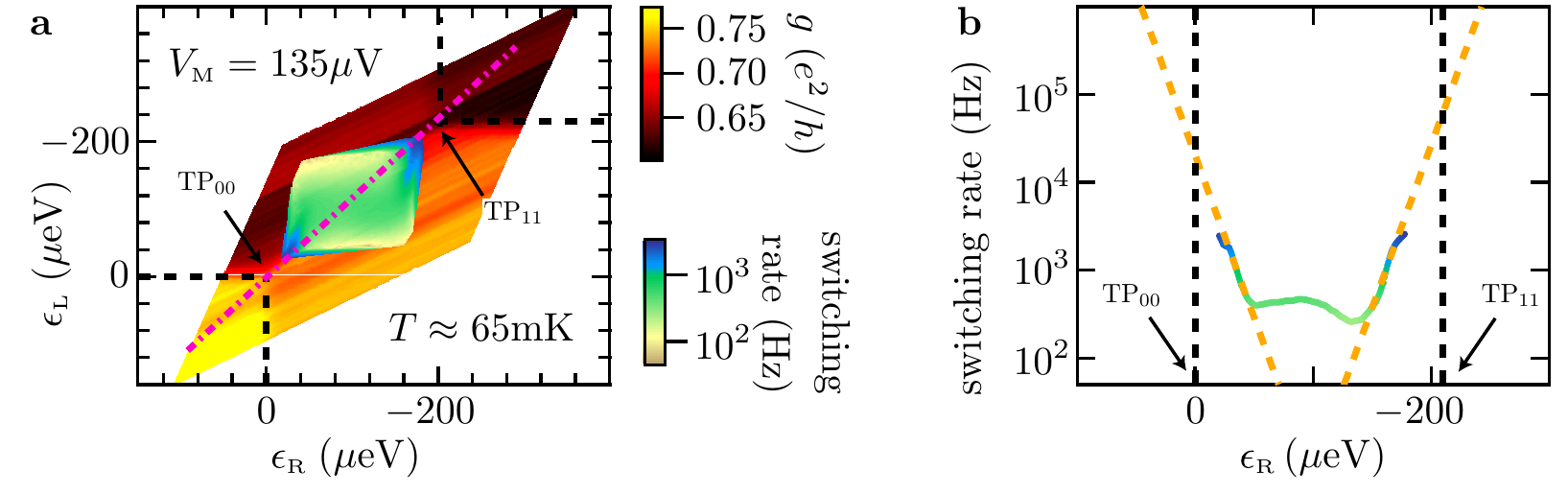}
	\caption{\textbf{Estimation of tunnel-coupling rates between the dots and their respective leads.} 
		\textbf{a}.  Charge stability diagram using the measured conductance through charge sensor (black-red-yellow colour scale) with the four main different values corresponding to the four double dot charge states, cf. Supplementary Fig.~\ref{fig:S3}. 
		The switching-rate is extracted by analysing individual real-time traces for each pixel and is overlaid on top of the conductance data (grey-green-blue colour scale), see~Ref.~\cite{Biesinger2014}. 
		The pink dashed line indicates the cut along the zero-detuning line shown in 
		\textbf{b}.  The range of switching-rate data available is limited by the finite bandwidth of the charge sensor.
		\textbf{b}. Cut of the switching-rate-map along the zero detuning line, cf.~pink dashed line in \textbf{a} (grey-green-blue coloured data points). 
		To estimate the bare tunnel rate of the quantum dots $\Gamma_{dd}$, the exponential region of the switching rate is fitted and extrapolated to the triple points TP$_{0,0}$ and TP$_{1,1}$, respectively (orange dashed curves). From this fit, we estimate $\Gamma_{dd}$ between \(10\) and \(200 \mathrm{ \mu V}\).
		The saturation  of the switching rate appearing between the exponential region ($\sim$ \(200\mathrm{ \mu V}\) to \(400\mathrm{ \mu V})\) is understood in terms of higher-order electron exchange effects (cotunnelling) via the leads~\cite{Biesinger2014} and is neglected in our model as it is very small.
		The switching rate flattens off for the highest rates measured due to finite sensor bandwidth. 
		This error-prone regime is ignored for the exponential fit.
	}
	\label{Fig:S4}
\end{figure*}

The charge stability diagrams are obtained by scanning the voltage on gate $V_\LL$ vs. $V_\RR$, see Supplementary Fig.~\ref{Fig:S1}\textbf{a}.
Upon changing these gate voltages, a linear feedback is applied to the sensor plunger gate to compensate for capacitive crosstalk with the charge sensor.
This correction bears no relevance to the effects discussed in this work and simply keeps the detector in a sensitive configuration.

Using an affine transformation
\begin{align}
	\begin{pmatrix}
		\epsilon_\LL \\
		\epsilon_\RR 
	\end{pmatrix}
=
\begin{pmatrix}
	l_{\LL\LL} & l_{\LL\RR}\\
	l_{\RR\LL} & l_{\RR\RR}
\end{pmatrix}
\begin{pmatrix}
	V_\LL-V_\LL^0 \\
	V_\RR-V_\RR^0
\end{pmatrix}
\end{align}
the two dimensional maps in the voltages \(V_\LL\) and \(V_\RR\) are transformed into the basis of $\epsilon_\LL$ and $\epsilon_\RR$, i.e., the energy of the left and the right quantum dot, respectively, see Supplementary Fig.~\ref{Fig:S1}\textbf{b}. 
The voltage offsets \(V_\LL^0\) and \(V_\RR^0\), are introduced such that the triple point \(\mathrm{TP}_{00}\) associated with the empty state occurs at \(\epsilon_\LL=\epsilon_\RR=0\).
The lever arms, \(l_{ij}\) have units of charge and quantify the energy shift in the \(i=\mathrm{L,R}\)  level due to a change in the \(j =\mathrm{L,R}\) gate voltage.
The relative magnitudes of the lever arms are found by ensuring that: (i) the (1,0)-(0,1) degeneracy line lies along \(\epsilon_\LL=\epsilon_\RR\), (ii) the degeneracy lines that involve only a transition in the left (right) dot are horizontal (vertical).
We calibrated the lever arm by fitting a Fermi-Dirac function to the lead transitions at elevated temperatures \cite{Maradan2014,Biesinger2014}.

The effect of sensor-dot backaction results in a deviation of the (1,0)-(0,1) charge degeneracy line from the conventional \textit{equilibrium phase boundary} of a standard double quantum dot, see Figs.~\ref{fig:setup}\textbf{a},\textbf{b}. 
To quantify this backaction effect, the energy difference between the measured and the conventionally-expected (1,0)-(0,1) degeneracy line is extracted along its full extent, i.e., between the two triple points where a degeneracy occurs also with the (0,0) or the (1,1) states, see Supplementary Fig.~\ref{Fig:S1}\textbf{c}. 
The maximal amplitude of this deviation, $\delta_\back$, is obtained by fitting a sine model to this extracted data, see Supplementary Information~\ref{SupNoteExp:3}. 
This amplitude $\delta_\back$ is extracted for different biases over the sensor quantum dot and temperatures, see Fig.~\ref{fig:temperature}.

While not discussed in the main article, we notice a difference in the magnitude between two amplitudes of the ``S''-feature: the deviation of the (0,1) charge state into conventional (1,0) region generally shows a smaller amplitude compared to deviation into the (0,1) region, see Supplementary Fig.~\ref{Fig:S1}\textbf{c}. This asymmetry is further discussed in Supplementary Information~\ref{SupNoteExp:4} and~\ref{spinassym}.
There we show that the spin degeneracy of the electronic levels can explain this asymmetry.
In the main text, $\delta_\back$ is extracted from the amplitude of the sine fit without taking the asymmetry of the ``S''-feature into account. 
As a consequence, $\delta_\back$ of the main text is the average of the two amplitudes.

\subsection{Quantifying $\delta_\back$ of ``S''-shape}\label{SupNoteExp:3}
We extract the contour curve at which the (1,0) and (0,1) charge states have equivalent probability (dashed curve in Supplementary Fig.~\ref{Fig:S1}\textbf{b}). 
For a system without any quantum-sensor backaction, this equi-probability line coincides with the zero-detuning (\(\epsilon_\LL = \epsilon_\RR\)) line. 
Here, however, we observe a measurement-induced ``S''-shaped deviation. 
In Supplementary Fig.~\ref{Fig:S1}\textbf{c}, we show a plot of the extracted difference between the zero-detuning line to the extracted ``S''-shaped contour curve. 
Such plots are used to quantify the amplitudes of the ``S''-shaped feature. 
We then obtain $\delta\back$, the maximal amplitude of the ``S''-shape feature, by fitting the extracted data with a sine model. 
Note that there is an offset present in the data, due to the asymmetry of the ``S''-feature discussed in Supplementary Information~\ref{SupNoteExp:4}, which we ignore in the main article.
We obtain the error bars in $\delta_\back$ in Fig.~\ref{fig:temperature} by estimating the uncertainty when extracting the ``S''-feature in the occupation imbalance map: 
Therefore, we compare $\delta_s$ for $\Delta=0$ (charge-degeneracy) with $\delta_\back$ for $\Delta = 0.2$ and $\Delta = -0.2$.
This corresponds to a 10\% uncertainty of the ratio $P_{(1,0)}/P_{(0,1)}$.

\subsection{Asymmetry of the ``S''-shape}
\label{SupNoteExp:4}

\begin{figure*}
	\centering
	\includegraphics[scale=1]{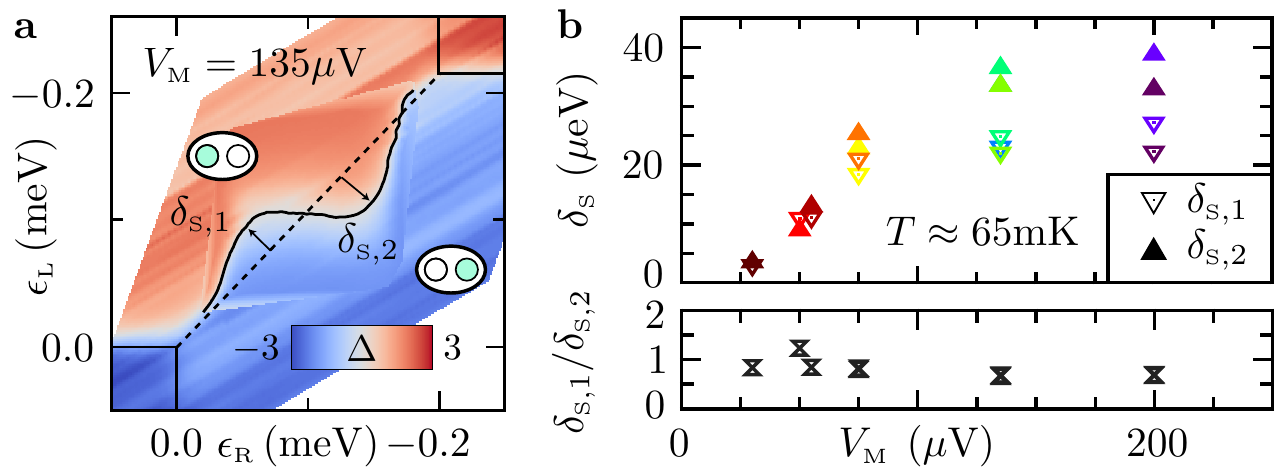}
	\caption{
		\textbf{Asymmetry in the backaction induced ``S'' shape.}
		\textbf{a}. Charge stability diagram taken at a sensor bias $V_M = 135\mathrm{ \mu V}$ which exhibits a pronounced asymmetry of the two amplitudes of the ``S''-features labelled $\delta_{\back,1}$ and $\delta_{\back,2}$.
		\textbf{b}. Dependence on the sensor bias voltage $V_M$ of the two amplitudes $\delta_{S,1}$ and $\delta_{\back,2}$. Each measurement is shown with an individual colour individual colour with triangles pointing upwards (downwards) for the amplitudes $\delta_{\back,1}$ ($\delta_{\back,2}$). The ratio $\delta_{\back,1}/\delta_{\back,2}$ is plotted on the right axis of the graph and remains about constant.
	}
	\label{Fig:S2}
\end{figure*}

We find an asymmetry of the ``S''-shape for larger sensor-bias voltages \(V_\MM\): the amplitude of the deviation from the zero-detuning line (\(\epsilon_\LL=\epsilon_\RR\)) closer to (0,0), labelled $\delta_{\back,1}$ in Supplementary Fig.~\ref{Fig:S2}\textbf{a}, is smaller than the deviation $\delta_{\back,2}$ located closer to the (1,1) state.
In Supplementary Fig.~\ref{Fig:S2}\textbf{b}, the triangles represent $\delta_{\back,2}$ (filled upwards-triangles) and $\delta_{\back,1}$ (empty downwards-triangles) for the individual measurements of the bias dependence. 
Individual measurements are presented in different colours. 
A clear trend is recognized in the data: a stronger quantum measurement backaction (larger sensor-bias voltage) leads to a larger absolute difference of the two amplitudes. 
The ratio $\delta_{\back,1}/\delta_{\back,2}\approx 0.8$, however, plotted as crosses in Supplementary Fig.~\ref{Fig:S2}\textbf{b} remains approximately constant. 
The outlier at \(V_\MM = 50\mathrm{\mu V}\) is attributed to difficulties in obtaining the amplitudes $\delta_{\back,1}$ and $\delta_{\back,2}$ in this regime, where almost no deviation from the zero-detuning line appears.
Note, that in Fig.~\ref{fig:temperature}, the average values of the individual measurements of Supplementary Fig.~\ref{Fig:S2}\textbf{b} are shown. 
As discussed in Supplementary Information~\ref{sup:tails}, this asymmetry clearly indicates that the particle-hole symmetry of the system is broken, likely due to the spin degree of freedom.

\subsection{Transition width}

To better understand the nature of the measurement backaction transition which we have observed, we investigate the scaling properties of the transition width.
First, we extract $\delta_\gamma$, the broadening of the (1,0)-(0,1) transition, by analysing cuts through the transition in the charge stability diagram (see Supplementary Fig.~\ref{Fig:S1}\textbf{b}). 
We present an example of such a cut in Supplementary Fig.~\ref{Fig:S4width}\textbf{a}. 
This data indicates a typical broadening in $\Delta$ of the transition from the charge state (1,0) to (0,1). 
To obtain $\delta_\gamma$, we fit this data with a scaled Fermi-Dirac function
\begin{align}
F(\epsilon,\delta_\gamma )= 4[1 + e^{(\epsilon-\lambda)/\delta_\gamma}]^{-1}-2,
\end{align}
where \(\lambda\) is an irrelevant free parameter that shifts the distributio left or right.
We then repeat this procedure for several cuts across the transition and average the result, before repeating the procedure for each charge stability diagram associated with a data-point in Supplementary Fig.~\ref{Fig:S4width}.

We find no dependence of $\delta_\gamma$ on the sensor bias $V_\MM$ as presented in Supplementary Fig.~\ref{Fig:S4width}\textbf{b}, which indicates that the measurement backaction does not dominate the broadening whereas it dominates the amplitude \(\delta_\back\). 
The data in Supplementary Fig.~\ref{Fig:S4width}\textbf{b} was obtained at a base-temperature of the dilution refrigerator, corresponding to an electronic temperature $T\approx 65 \mathrm{ mK}$ (\(0.026U/k_\BB\)).
Next, we investigate the temperature dependence of $\delta_\gamma$ at a fixed sensor bias voltage $V_\MM = 75\mathrm{ \mu V}$ (\(0.35U/e\)), see Supplementary Fig.~\ref{Fig:S4width}\textbf{c}. 
When we increase the temperature, we find a linear dependence of the broadening $\delta_\gamma$ upon increasing the electron temperature $T$. 
Furthermore, we find that the transition width \(\delta_\gamma\) as a function of temperature \(T\) is in good agreement with the thermal energy $\delta_\gamma \approx k_\BB T$~\cite{DiCarlo2004}.

As the width shrinks linearly with diminishing temperature, the experiment is in agreement with a phase transition.
A saturation of the width at lower temperatures \(\delta_\gamma(T\to 0)>0\) is still possible and would indicate the presence of an abrupt crossover, with a small but finite width~\cite{goldstein2010population}. 
The transition width is strongly tied to the nature of the tails, see Supplementary Information~\ref{sup:tails}, motivating further studies to investigate the microscopic details of the broadening mechanism.

\begin{figure*}
	\includegraphics[scale=1]{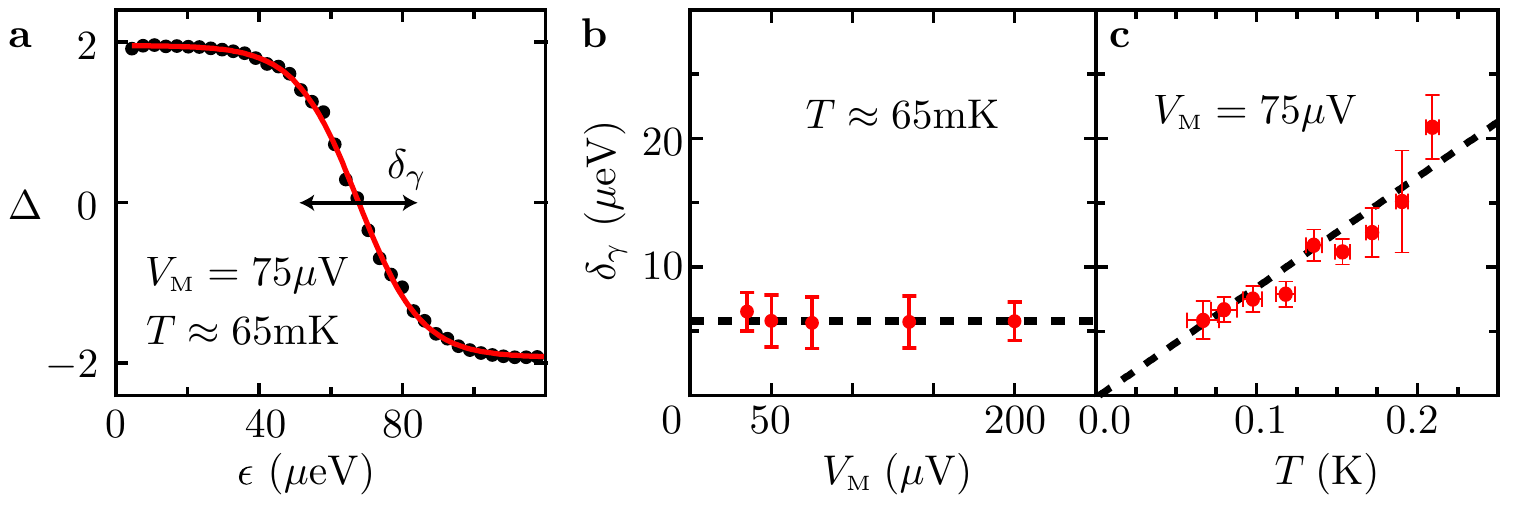}
	\caption{\label{Fig:S4width}
		\textbf{Transition width.} 
		\textbf{a}.
		Population imbalance $\Delta$ along a cut perpendicular to the (1,0)-(0,1) transition in the charge stability diagram, such as the one in Supplementary Fig.~\ref{Fig:S1}. Here, \(\epsilon\) is the Euclidean distance along the cut. We obtain the interdot transition broadening $\delta_\gamma$ by fitting the data to a Fermi-Dirac distribution.
		For each charge stability diagram, we average the width over three cuts in the energy basis (\(\epsilon\)) and one cut in the voltage basis~(\( V\)).
		\textbf{b}.
		The width $\delta_\gamma$ shows no clear dependence on the sensor bias voltage $V_M$ and, therefore, does not depend on the backaction. The dashed line indicates the thermal energy~\(k_\BB T\) associated with the electronic temperature \(T=65\)mK.
		\textbf{c}.
		Temperature dependence of the width $\delta_\gamma$ at fixed measurement bias voltage $V_M = 75 \mathrm{ \mu V}$.
		Over the entire temperature range in our experiments, the extracted width is similar to the thermal energy $k_\BB T$ (dashed black line).
	}
	\label{fig:width}
\end{figure*}

\subsection{Estimation of $\alpha$}\label{SupNoteExp:2}

We now describe the estimation of the ratio between the Coulomb interaction of sensor-to-left dot ($U_{LM}$) and the sensor-to-right dot ($U_{RM}$): $\alpha^2 = U_{LM}/U_{RM}$. Starting from an empty DD charge configuration, (0,0), the change in sensor conductance $g_\Delta $ when adding an electron to the left dot $| g^{(0,0)} - g^{(1,0)}|$ is different from that observed when adding an electron to the right dot $| g^{(0,0)} - g^{(0,1)}|$. Under the assumption that, in the region of interest, the sensor conductance is linear in sensor quantum dot energy, we obtain an approximation of $\alpha$ by comparing the magnitudes of $g_\Delta $ for different charge transitions, see Supplementary Fig.~\ref{fig:S3}.

Generally, we observe four main values of conductance $g$ through the charge sensor, corresponding to the four relevant charge configurations of the DD. This is demonstrated in Supplementary Fig.~\ref{fig:S3}: here a histogram of the charge sensor signal for a charge stability diagram around the (1,0)-(0,1) double dot transition is presented. From the specific data shown in Supplementary Fig.~\ref{fig:S3}, we obtain $\alpha^2 = U_{LM}/U_{RM} \approx ( g^{(0,0)} - g^{(1,0)}) /  ( g^{(0,0)} - g^{(0,1)}) = 5.4$.
Repeating the process for several distinct experimental parameters we obtain an estimate \(\alpha = 2.2\pm 0.2\) which we use in the theoretical model.

\begin{figure*}
	\includegraphics[scale=1]{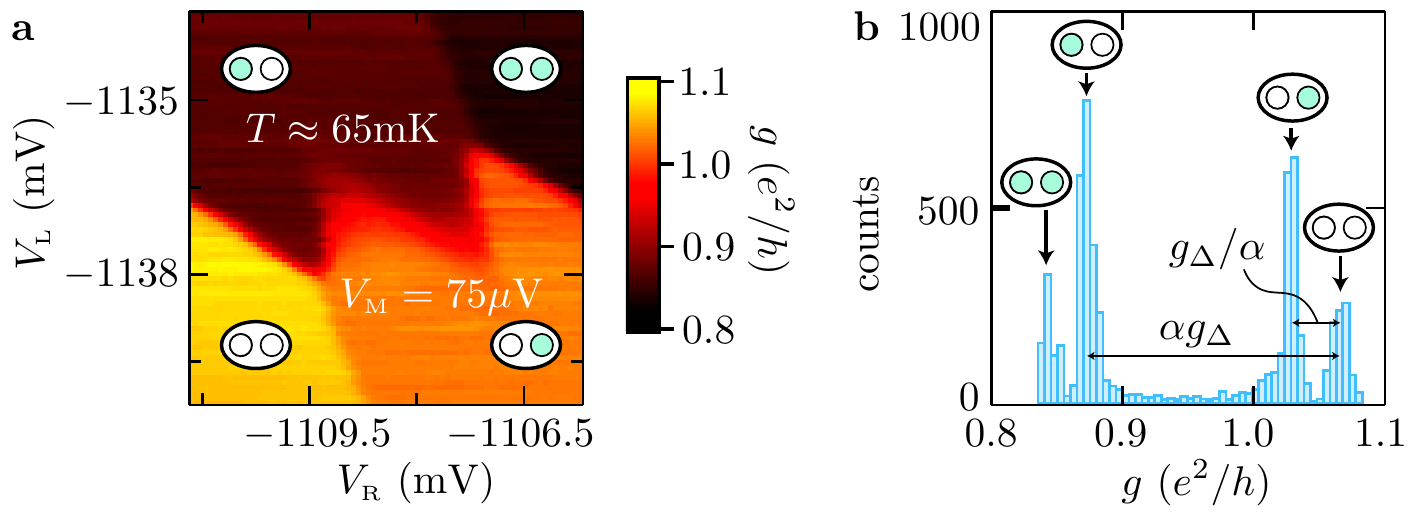}
	\caption{\textbf{Measuring the asymmetry \(\alpha\) of the level widths.} \textbf{a}. Sensor conductance $g$ of the charge stability diagram around the $(1,0)$-$(0,1)$  charge state configurations of the double dot. Due to different capacitive coupling of the DD to the charge sensor, the four charge states are distinguishable by their characteristic sensor conductances. \textbf{b} Histogram of the charge stability diagram shown in (a). Each charge state in panel (a) appears at different but roughly constant sensor conductance which results in peaks when plotted as a histogram. The parameter $\alpha$ is estimated by the ratio of the difference $g_\Delta$ in conductance $g$ of the charge states involved.
	}
	\label{fig:S3}
\end{figure*}

\section{Model}
\subsection{Microscopic model}

The full microscopic model describing the system--detector setup can be written using the Hamiltonian
\begin{align}
H = H_{\doubledot} +H_{\leads} + H_{\tunnel} + H_{\MM} + H_{\interaction},
\end{align}
including the double quantum dot $H_{\doubledot}$, its leads $H_{\leads}$, tunnel-coupling between the dots and their respective leads $H_{\tunnel}$, the detector model $H_{\MM}$, and the system--detector interaction $H_{\interaction}$ terms. 
The double dot Hamiltonian \(H_\doubledot = H_\LL+H_\RR+H_{\mathrm{coupl}}\) is in turn described by the Hamiltonians \(H_\LL\) and \(H_\RR\) of the left and righ dots, and a coupling Hamiltonian \(H_{\mathrm{coupl}}\).
Assuming at most a single orbital mode in each of the dots and in the absence of magnetic field, we write the left dot Hamiltonian as
\begin{align}
	H_\LL = \sum_\sigma \epsilon_\LL d^\dagger_{\LL\sigma}d^\pdagger_{\LL\sigma}
	+U_\LL d^\dagger_{\LL\uparrow}d^\pdagger_{\LL\uparrow}d^\dagger_{\LL\downarrow}d^\pdagger_{\LL\downarrow}
\end{align}
where $d_{\LL\sigma}^\dagger$ ($d_{\LL\sigma}^\pdagger$) creates (annihilates) an electron with spin \(\sigma\) and energy $\epsilon_\LL$ in the left dot.
The onsite Coulomb repulsion~\(U_\LL\) is much larger than all relevant energy scales in the system such that the doubly occupied state of the left dot is energetically forbidden.
Due to the absence of magnetic field the two spin states are degenerate, such that in our analysis the difference between including and excluding spin is the inclusion of a degeneracy factor, see Supplementary Information~\ref{sup:spin}.
The right dot Hamiltonian is obtained in an analogous way to \(H_\LL\) but with the substitution \(\mathrm{L} \to \mathrm{R}\).
In the following we drop the spin degree of freedom and associated index \(\sigma\) for simplicity.
The left and right dot Hamiltonians then become
\begin{align}
	H_i = \epsilon_i^\pdagger d_i^\dagger d_i^\pdagger
\end{align}
for the left $i=\mathrm{L}$ and right $i=\mathrm{R}$ dots.
The two dots are only electrostatically coupled, as the tunnelling barrier between them is very large, such that the coupling Hamiltonian becomes
\begin{align}
	H_{\mathrm{coupl}} = U d_\LL^\dagger d_\LL^\pdagger d_\RR^\dagger d_\RR^\pdagger,
\end{align}
where the mutual charging energy (experimentally estimated $U\approx215 \mathrm{\mu eV}$) penalises the simultaneous occupancy of the left and right dots.
The total double dot Hamiltonian is thus
\begin{align}
H_\doubledot =
\epsilon_\LL^\pdagger d_\LL^\dagger d_\LL^\pdagger
+
\epsilon_\RR^\pdagger d_\RR^\dagger d_\RR^\pdagger
+
U d_\LL^\dagger d_\LL^\pdagger d_\RR^\dagger d_\RR^\pdagger,
\end{align}
which has the empty, left-, right- and doubly-occupied states as eigenstates.

The left and right leads are described by
\begin{align}
H_\leads = \sum_{k,i=\LL,\RR} \epsilon_{i k}^\pdagger c_{ik}^\dagger c_{ik}^\pdagger,
\end{align}
where $k$ indexes the different momenta of the leads and $c_{ik}^\dagger$ ($c_{ik}^\pdagger$) creates (annihilates) an electron with energy $\epsilon_{ik}$ in the left $i=\mathrm{L}$ or right $i=\mathrm{R}$ leads.
Each of these leads is further associated with a tunable chemical potential $\mu_{\LL,\RR}$, which is kept constant and at equilibrium, serving as the energy reference for the experiment, $\mu_{\LL}=\mu_{\RR}=0$.
This is described by the tunnelling Hamiltonian
\begin{align}
H_\tunnel =  \sum_{k,i=\LL,\RR}
t\,( d_{i}^\dagger c_{ik}^\pdagger + h.c.),
\end{align}
where the tunnelling amplitude \(t\) is taken to be momentum independent and equal for both dots, as is the case in the experiment.

The detector Hamiltonian is built up in a similar fashion but with a single dot
\begin{align}
H_{\MM}= \epsilon_\MM^\pdagger d_\MM^\dagger d_\MM^\pdagger +\!\!\!\!\!
\sum_{k,i=\source,\drain}\!\big[ \epsilon_{i k}^\pdagger c_{ik}^\dagger c_{ik}^\pdagger \!+\!
t_\MM^\pdagger(d_{\MM}^\dagger c_{ik}^\pdagger \!+\! h.c.)\big].
\end{align}
Here all quantities are defined analogously to the double dot and its leads, but with new indices for the detector dot ($\mathrm{M}$), as well as for the source ($\source$) and the drain ($\drain$) detector leads.
A bias voltage across the detector $V_\MM$ is directly proportional to difference between source and drain chemical potential $\mu_{\source \drain} = \mu_\source-\mu_\drain$. Finally, the interaction Hamiltonian describes capacitive coupling between the measurement dot and both the left and right dots
\begin{align}
H_{int}=
(
U_{\LL \MM}^\pdagger d^\dagger_\LL d^\pdagger_\LL
+
U_{\RR \MM}^\pdagger d^\dagger_\RR d^\pdagger_\RR
)
d^\dagger_\MM d^\pdagger_\MM.
\end{align}
Here, we introduced the two Coulomb interaction strengths $U_{\LL \MM}^\pdagger$ and $U_{\RR \MM}^\pdagger$ between the left or right dot and the measurement dot.
Crucial to our work, these interaction terms are not equal due to the different distance between the detector and the two dots. We define their ratio as
\begin{align}
\alpha^2=U_{\LL \MM}/U_{\RR \MM}.
\end{align}
As shown in Supplementary Information~\ref{SupNoteExp:2}, this value is experimentally determined to be \(\alpha = 2.2\pm 0.2\).

\subsection{Rate equation with backaction}
We analyse transport through the DD using a rate equation. To this end, we assume a sufficiently small tunnel coupling between the dots and their leads $\rho_i
\left|t\right|^2\ll U$ with $\rho_i$ the density of states of lead $i$. We, then, perturbatively derive the corresponding sequential tunnelling rates using Fermi's golden rule~\cite{Bruus2004}.
In the absence of the detector, Fermi's golden rule for the symmetrically-coupled DD reads~\cite{Biesinger2014}
\begin{align}
\label{eq:rates}
\!\!\!\Gamma_{if} = \Gamma_{\doubledot}\!\!\int\!\! \dd {\epsilon} \delta(\epsilon-\epsilon_i+\epsilon_f)n_\FF(\epsilon)
=\Gamma_{\doubledot} n_\FF(\epsilon_f-\epsilon_i),
\end{align}
where $i,f$ denote initial and final states (described by different charge configurations $\left(N_\LL, N_\RR\right)$ of the DD), $\Gamma_\doubledot=2\pi \rho\left|t\right|^2/\hbar$ is the bare tunnelling rate with $\rho_i=\rho$, and we introduced the Fermi-Dirac distribution
\begin{align}
n_\FF(\epsilon) = \frac{1}{1+\exp(\epsilon/k_\BB T)}.
\end{align}

One of the central results of Ref.~\cite{Zilberberg2014} is that, when calculating such transport rates through the system, a continuous charge measurement of a quantum dot enters as an effective width for the dot's energy level. Using this result, we effectively trace out the detector and incorporate its impact directly in the a slightly modified expression for the rates
\begin{align}
\Gamma_{if}=\Gamma_{\doubledot} \int \dd{\epsilon}\frac{1}{\pi} \frac{\gamma_{f-i}}{\epsilon^2+\gamma_{f-i}^2}n_\FF(\epsilon_f-\epsilon_i+\epsilon),
\end{align}
where the width \(\gamma_{f-i}\) depends on the relevant dot level associated to the specific rate. For example, if $f=(1,1)$ and $i=(1,0)$ then \(\gamma_{f-i} = \gamma_\RR\) because only the right dot is involved in the tunnelling process.
Performing the integral we obtain
\begin{align}
\label{eq:broadRates}
\Gamma_{if}=\Gamma_{\doubledot}n_\LL(\epsilon_f-\epsilon_i,\gamma_{f-i}),
\end{align}
where we have introduced the modified Fermi-Dirac distribution
\begin{align}
n_\LL(\epsilon,\gamma) = \frac{1}{2} + \mathrm{Im}\psi\left(\frac{1}{2}+\frac{\gamma-i\epsilon}{2\pi k_\BB T}\right),
\end{align}
and \(\psi\) is the digamma function.
These rates have algebraic tails which decay as \(\sim \gamma/\epsilon\) for large \(\epsilon\)~\cite{Bruus2004}.
We use this phenomenological approach motivated by previous works~\cite{Zilberberg2014, Bischoff2015}, and leave more detailed calculations for future work.

We substitute the rates~\eqref{eq:broadRates} into a rate equation, describing the time evolution of the occupation probability of each charge state $i$
\begin{align}
\label{eq:rateEq}
\partial_t P_i = \sum_j P_j \Gamma_{ij} - P_i \sum_j\Gamma_{ij},
\end{align}
leading to a Markovian chain as illustrated in Fig.~\ref{fig:model}\textbf{a}. Solving for the steady-state $\partial_t P_i=0$, we obtain the mean/observable charge states of the DD and thus the imbalance \(\Delta\), see Supplementary Fig.~\ref{fig:theoryMap}. 
We can then extract \(\delta_\back\) from these charge stability diagrams and use the result to fit \(\chi\) and \(\xi\), see Fig.~\ref{fig:temperature}.

\begin{figure}
	\includegraphics[scale=1]{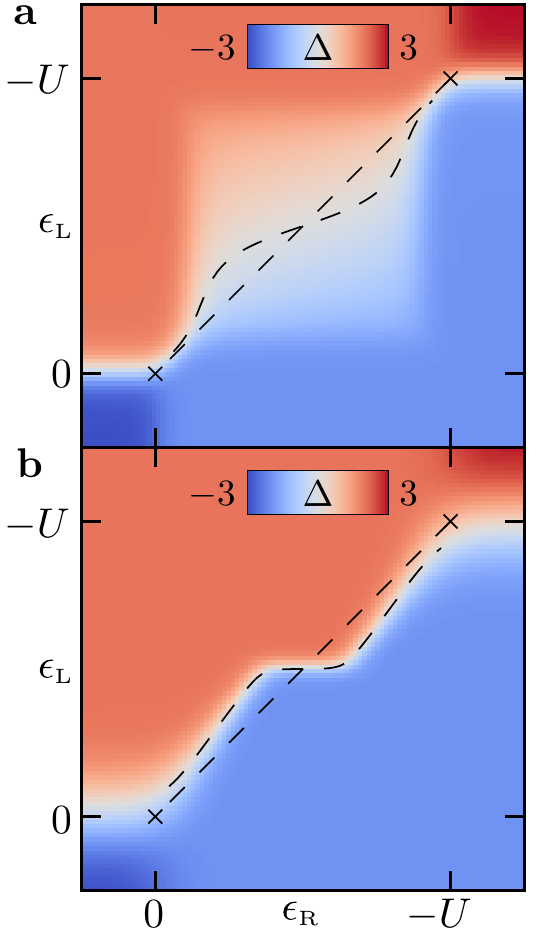}
	\caption{\label{fig:theoryMap}
		\textbf{Theoretically calculated imbalance $\Delta$ using the rate equation, while including the effect of measurement backaction.}
		We use the same parameters as in the experimental plot Fig.~\ref{fig:setup}\textbf{b}. The measurement bias is \(V_\MM = 0.35 U\), temperature is \(T = 0.026 U/k_\BB \) and the asymmetry in the coupling is \(\alpha = 2.2\).
		{\bf a}.
		Using the Lorentzian broadening~\eqref{eq:broadRates} of the DD due to backaction, with the fit parameters~\(\chi=9.4\times10^{-3}\),  and \(\xi =2.5\times 10^{-3}U\).
		{\bf b}.
		Using a Gaussian broadening~\eqref{eq:gaussWidth} of the DD due to backaction, with the fit parameters~\(\chi^\GG= 0.099\), and \(\xi^\GG = 0.084U\).
		We notice that the shape of the anomaly is more similar to the experiment in the Lorentzian case, while the width of the transition is better captured by the Gaussian approach.
	}
\end{figure}

\subsection{Nature of the tails}
\begin{figure*}
	\includegraphics[scale=1]{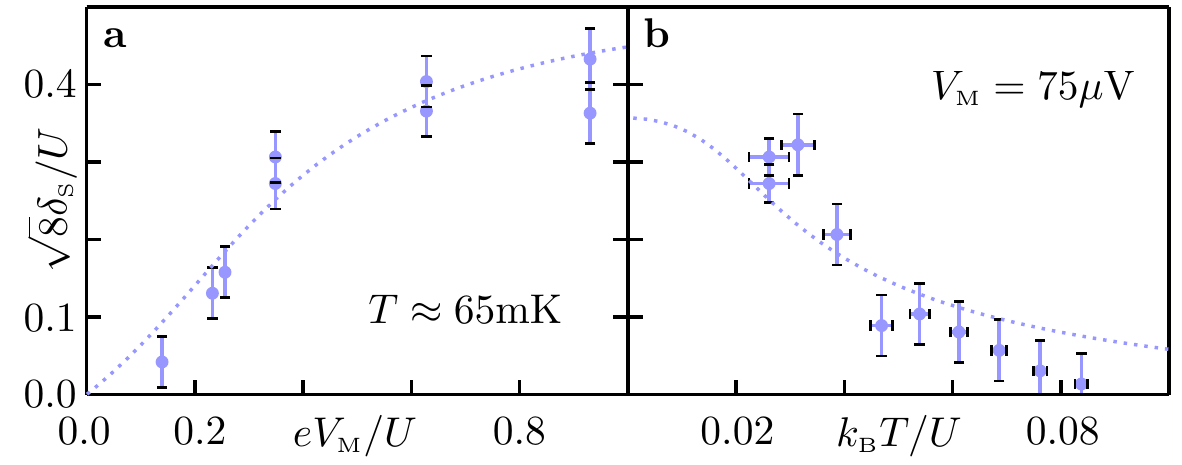}
	\caption{\label{fig:supTemp}
		\textbf{Fitting the imbalance with Gaussian broadenings.}
		Measured maximal deviation $\delta_{\back}$ (circles) as a function of the measurement bias voltage $V_{\MM}$ ({\bf a}) and temperature $T$ ({\bf b}) as in Fig.~\ref{fig:temperature}.
		A Gaussian broadening of the levels, leads to a qualitatively similar scaling (dotted line) of the maximal deviation~\(\delta_\back\) as both the experiment and the Lorentzian broadening, cf.~Fig.~\ref{fig:temperature}.
		Unlike the Lorentzian broadening, there is no clear saturation for a given \(\alpha\) in the case of Gaussian broadening.
	}
\end{figure*}

\label{sup:tails}
In Ref~\cite{Zilberberg2014}, backaction was predicted to cause a Lorentzian broadening of the delta function in Fermi's golden rule~\eqref{eq:rates}, leading to the broadened rates~\eqref{eq:broadRates}.
Here, we point out that the shape of the tails has a strong influence on the exact form of the population switching region, or ``S''-shape, and the width of the transition.
We use a Gaussian widening of the level as a counterweight to the Lorentzian ones.
While there are no works suggesting such a broadening, it is useful as an extreme example, which is completely opposed to the Lorentzian.
These distributions have very different properties, which manifest in the shape and contrast of the charge stability diagrams, see Supplementary Fig.~\ref{fig:theoryMap}.
However, the qualitative behaviour of the population switching and the scalings with temperature and bias, remain the same, see Supplementary Fig.~\ref{fig:supTemp}.
This motivates future work, which will investigate the nature of the population switching transition at low temperatures, to determine among other things whether this is a phase transition or abrupt crossover.
A likely contribution to the specific shapes is also that the Lorentzian widths~\(\gamma\) may in general depend on the parameters \(\epsilon_\LL,\epsilon_\RR,T\) of the system.

To include a Gaussian width we replace the delta function~\(\delta\) in the rate calculation~\eqref{eq:rates} through
\begin{align}
\label{eq:gaussWidth}
\delta(\epsilon-\epsilon_i+\epsilon_f) \to
\frac{1}{\sqrt{\pi}\gamma_{f-i}} \exp \left[\frac{-(\epsilon-\epsilon_i+\epsilon_f)^2}{\gamma_{f-i}^2}\right],
\end{align}
and then evaluate the integral numerically.
Recall that \(\gamma_{f-i}\) is either \(\gamma^\GG_\LL\) or \(\gamma^\GG_\RR\) depending on which level is involved in the transition \(i\to f\) (We use \(\mathrm{G}\) to indicate parameters that belong to the Gaussian model).
We can then use the rate equation~\eqref{eq:rateEq}, to compute the steady-state by imposing \( \partial_t P = 0 \), both for the Lorentzian and Gaussian broadenings, see Supplementary Fig.~\ref{fig:theoryMap}.
From there, we extract a width \(\delta_{\back}\) for a given set of fitting parameters \(\chi\), \(\xi\) and the experimental parameters~\(V_\MM,T\) and \(\alpha\).
In the main text, for the Lorentzian we used only the largest of the background or bias induced widths $\gamma_{j} = \max(\xi, \gamma_{\back, j})$, which is functionally similar to root square addition~\(\gamma_{j} = \sqrt{\xi^2+ \gamma^2_{\back, j}}\) of the widths, which is typical of Lorentzian line broadening.
However such an approach to the Gaussian widths cannot be expected to capture the experimentally observed signatures.
Specifically, the growth of one of the widths while the other remains constant causes the rapid formation of a very large \(\delta_\back\), due to the rapid decay in the tails of the Gaussian.
Instead of simply using the larger of the two rates, we therefore add the two contributions linearly for the Gaussian, such that
\begin{align}
\gamma^{\GG}_{j} = \xi^\GG + \gamma^\GG_{\back, j},
\end{align}
where~\(\gamma^\GG_{\back, \LL} = \alpha\chi^\GG V_\MM\), and ~\(\gamma^\GG_{\back, \LL} = \chi^\GG V_\MM/\alpha\).
Of course a microscopic investigation of the exact nature of the tails must also provide a prescriptive way of adding the widths.
Here however, we aim to qualitatively understand the physical processes at stake and thus leave these details to future works. 
We calculate the imbalance \(\Delta\) as a function of \(T\) and \(V_\MM\) to fit  \(\chi^\GG = 0.099\) and \(\xi^\GG = 0.084 U\), see Supplementary Fig.~\ref{fig:supTemp}.
Note, that the resulting value for \(\xi\) is nearly an order of magnitude larger than the expected value~\(\sim 1 \mathrm{\mu e V}\)~\cite{Camenzind2018HyperfinephononSpinRelaxation}.

In a typical experimentally relevant situation, cf. Fig~\ref{fig:setup}\textbf{b} and~\ref{fig:theoryMap}, we notice that shape of the ``S'' feature is better captured by the Lorentizian broadening, while the width is better described by the Gaussian. 
We conclude that the precise form of the broadening can be investigated using the complete charge stability diagrams.
While our brief description here suggests an intermediate \textit{between} a Lorentzian and a Gaussian, this can be achieved in a multitude of ways, e.g. parameter (\(\epsilon\)) dependent broadenings (\(\gamma\)) or different power law decays.
Furthermore, as (part of) the CSD level lies in the measurement bias window we expect resonant effects, which must be resummed to be accounted for properly.
A detailed future study of the sub-splitting of region (ii) in Fig.~\ref{fig:model}\textbf{c} will have to take such effects into account.

\subsection{Asymmetry in the ``S'' -shape}
\label{spinassym}
\begin{figure}
	\includegraphics[scale=1]{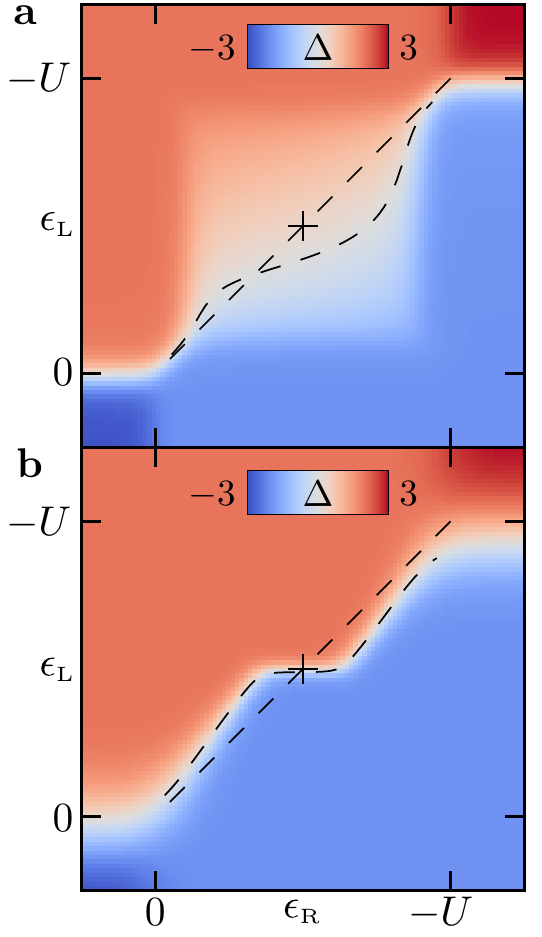}
	\caption{\label{Fig:supmatSpin}
		\textbf{Effect of spin.}
		A theoretical prediction for the ``S''-shape for a model with spin degenerate levels, for both Lorentzian~(\textbf{a}) and Gaussian~(\textbf{b}) broadenings.
		In the Lorentzian case we notice a significant particle-hole symmetry breaking, while in the Gaussian case it is barely visible (distance between the ``S''-shape and the black cross)
		Parameters as in Fig.~\ref{fig:theoryMap}, except for the inclusion of spin, see Eq.~\eqref{eq:spin}.
	}
\end{figure}
\label{sup:spin}

As discussed in Supplementary Information~\ref{SupNoteExp:4}, the experimental data shows an asymmetry in the ``S''-shape, which breaks the particle-hole symmetry of the~\(\epsilon_\LL=-U/2\), \(\epsilon_\RR=-U/2\) configuration.
If both left and right levels are allowed to be spin degenerate (as expected in the zero magnetic field experiment) this leads to overall degeneracies of \(1,2,2,4\) for the empty, left-, right- and doubly-occupied states respectively.
The fact that the empty and doubly-occupied states have different degeneracies manifestly breaks particle-hole symmetry.
We conclude that spin degeneracy is a candidate for the asymmetry seen in the experiment.
To further investigate this property, we include the degeneracies in our rate equation, which leads to a transformation
\begin{align}
\label{eq:spin}
\Gamma^+\to 2\Gamma^+, \quad \Gamma^-\to\Gamma^-.
\end{align}
in the rates which add or remove electrons from the DD.

The result of performing this substitution is very different in the case of Lorentzian broadening when compared to  Gaussian broadening, see Supplementary Fig.~\ref{Fig:supmatSpin}\textbf{a} and~\textbf{b} respectively.
In the latter case, the decay of the rates as a function of \(\epsilon_\LL\), and \(\epsilon_\RR\) is exponential.
Thus as the widths are relatively small the factor of two in the rates causes only a small shift in the intersection between the diagonal and the ``S''-shape.
On the other hand, the tails of the Lorentzian are algebraic and do, therefore, not have a characteristic scale on which they decay. 
The intersection between the diagonal and the ``S''-shape can thus shift significantly even for small broadenings.

We conclude that spin is a likely candidate for the experimentally observed particle-hole asymmetry, see Supplementary Information~\ref{SupNoteExp:4}.
Furthermore, the very large (small) asymmetry in the Lorentzian (Gaussian) case compared to the experimental evidence in Supplementary Fig.~\ref{Fig:S2} again indicates that the broadening is sharper than Lorentzian.
This strong dependence of the asymmetry on the type of tails, see Supplementary Fig.~\ref{Fig:supmatSpin}, shows that it can be used as a signature to investigate the precise nature of the tails.

\end{document}